\documentclass[twocolumn,amsmath,amssymb,prb]{revtex4}

\usepackage{graphicx}
\usepackage{dcolumn}
\usepackage{bm}
\usepackage{pstricks}

\begin{document}

\preprint{PRL}

\title{Electron-hole spin flip-flop in semiconductor quantum dots}

\author{Y. Benny}
\email{byael@tx.technion.ac.il}\affiliation{The Physics Department
and the Solid State Institute, Technion -- Israel Institute of
Technology, Haifa 32000, Israel.}
\author{R. Presman}\affiliation{The Physics Department
and the Solid State Institute, Technion -- Israel Institute of
Technology, Haifa 32000, Israel.}
\author{Y.Kodriano}\affiliation{The Physics Department
and the Solid State Institute, Technion -- Israel Institute of
Technology, Haifa 32000, Israel.}
\author{E. Poem}\affiliation{The Physics Department
and the Solid State Institute, Technion -- Israel Institute of
Technology, Haifa 32000, Israel.}
\author{T. A. Truong}\affiliation{Materials Department, University of California, Santa Barbara, California 93106, USA.}
\author{P. M. Petroff}\affiliation{Materials Department, University of California, Santa Barbara, California 93106, USA.}
\author{D. Gershoni}
\affiliation{The Physics Department and the Solid State Institute, Technion -- Israel Institute of Technology, Haifa 32000, Israel.}

\date{\today}

\begin{abstract}
We use temporally resolved intensity cross-correlation measurements
to identify the biexciton-exciton radiative cascades in a negatively
charged QD. The polarization sensitive correlation measurements show
unambiguously that the excited two electron triplet states relax
non-radiatively to their singlet ground state via a spin non
conserving flip-flop with the ground state heavy hole. We explain
this mechanism in terms of resonant coupling between the confined
electron states and an LO phonon. This resonant interaction together
with the electron-hole exchange interaction provides an efficient
mechanism for this, otherwise spin-blockaded, electronic relaxation.
\end{abstract}

\pacs{Valid PACS appear here}
\maketitle Semiconductor quantum dots (QDs) have received considerable
attention over the years, due to their atomic-like features and
their compatibility with modern microelectronics. QDs are
particularly attractive as the key ingredients in bright solid-state sources of single and entangled photons,~\cite{Akopian06} and as
excellent interfaces between photons~\cite{Turchette95,Knill00} (flying qubits) and
confined charge carriers spins~\cite{Kosaka08,Benny10} (anchored matter qubits).  The
spins of QD confined charge carriers are promising candidates for
implementations of qubits and quantum gate
operations.~\cite{Kosaka09,Press10,Benny10,Kodriano12} Indeed,
coherent control of confined carriers' spins have been studied and
demonstrated in various experimental
ways.~\cite{Kosaka09,Press10,Kodriano12} Studies of spin dephasing
and decay in general, and the controlled preparation of multi-carrier
spin states, in particular, remain important challenges.

Here, we experimentally identify and theoretically explain a QD relaxation mechanism involving a deterministic electron-heavy-hole spin flip-flop.
For the observation of this effect we use temporally resolved
polarization sensitive intensity correlation measurements of
two-photon radiative cascades that resulted from sequential
recombination of QD confined electron-hole pairs in the presence of
an additional electron. An efficient non-conserving spin decay
process between the excited triplet state and
its ground singlet state is observed. During this relaxation, the
total spin projection of the two electrons on the QD symmetry axis
changes from unity to zero. We show that this happens only with an accompanying
spin flip of the ground state heavy hole. We explain this spin
flip-flop as resulting from the Fr\"{o}hlich interaction between the
electrons and an LO phonon together with the electron-hole exchange
interaction. Though strong, quasi-resonant electron-LO phonon
interaction in single QDs was reported and modeled
previously,~\cite{Hameau01,Hameau99,Hameau02,Stauber00,Kaczmarkiewicz10,Jacak02,Melnikov01} here, we show for the first time that it provides a fast and efficient, almost deterministic,
non-conserving spin decay mechanism. For completeness, we use our model for calculating the electronic relaxation in the absence of the additional electron, and compare it with spectral measurements of a neutral QD.

The sample that we study was grown by molecular-beam epitaxy on a
(001) -oriented GaAs substrate. One layer of strain-induced InGaAs
QDs was placed in a one wavelength microcavity formed by two distributed Bragg reflecting
mirrors. The microcavity was optimized for the range of wavelengths in which the QDs emit photoluminescence (PL). The measurements were carried out in a micro-PL setup at 4.2 K. The setup provides spatial resolution of about $1\mu $m, spectral
resolution of about 10 $\mu$eV and temporal resolution of about 400 ps in measuring the arrival times of two photons originating from two different spectral lines, at given polarizations. More details about the sample~\cite{Garcia97, DekelSSC} and the micro-PL setup~\cite{Poem10,Benny11} are given in earlier publications.

\begin{figure}
\includegraphics[width=0.48\textwidth]{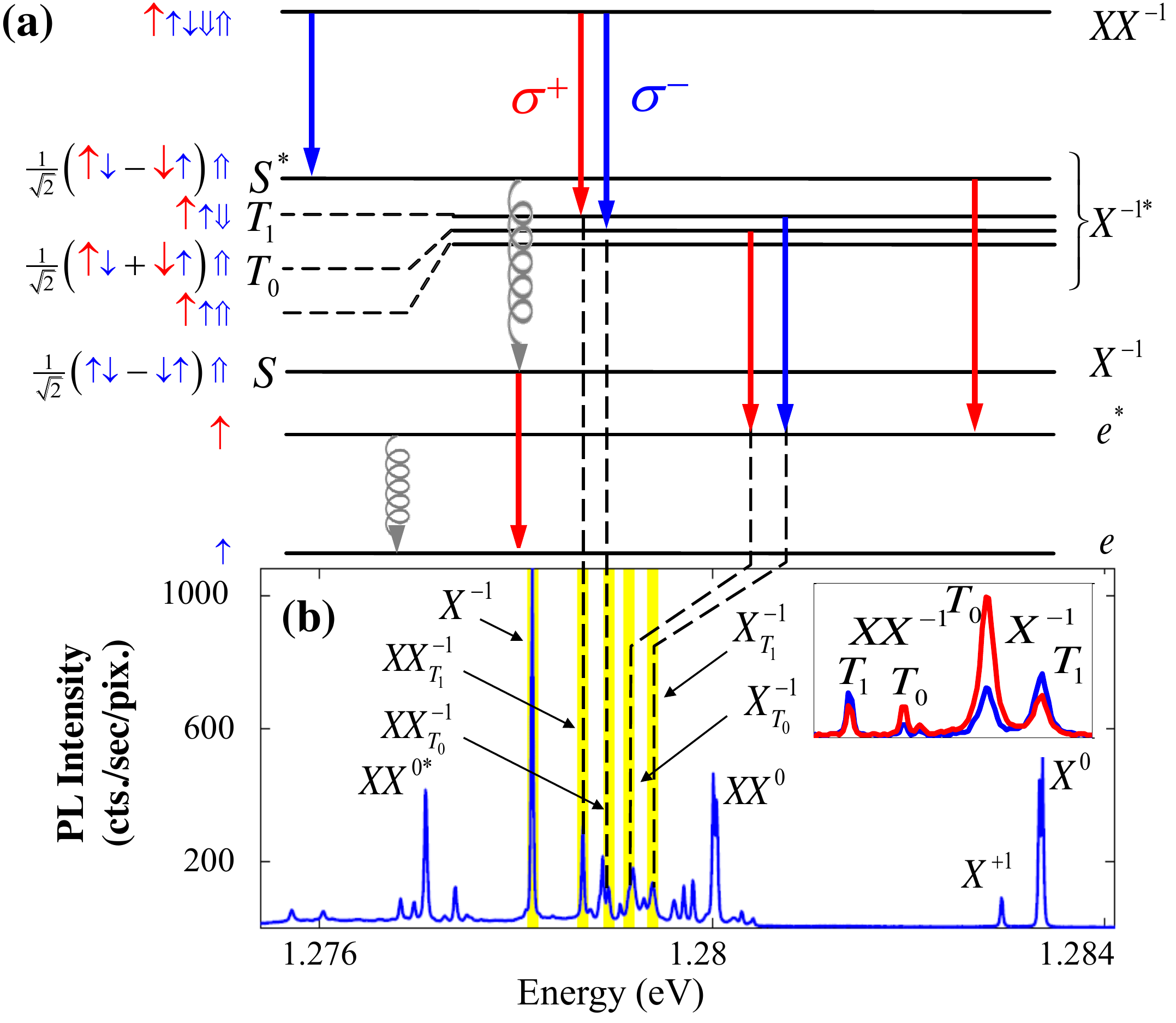}
\caption{\label{fig:EnergyDiagram} (a) Energy levels diagram of an
optically excited, singly negatively charged QD (each level is
doubly, Krammers' degenerate). The optical transitions between these
levels are marked by vertical arrows where blue (red) arrows
represent left (right) hand circular polarization. Curly arrows
represent nonradiative spin conserving phonon assisted relaxations.
The spin configurations and the total spin projection on the QD
symmetry axis  are presented to the left of each level. Blue (red)
arrows represent ground (excited) states and single (double) arrow
represents electron (hole) spin. (b) Measured PL spectrum of a
singly negatively charged QD. The observed spectral lines are
conventionally marked and linked to the optical transitions in (a) by vertical dashed lines. The inset presents horizontal (blue) and
vertical (red) linearly-polarized PL spectrum of the $T_0$ and $T_1$
excitonic (right doublet) and biexcitonic (left doublet) transitions.}
\end{figure}

Figure \ref{fig:EnergyDiagram}(a) presents an energy level diagram
of a singly negatively charged QD, optically excited with two
excitons (biexciton). For simplicity, only the state with a spin up unpaired electron is described, but all the levels are doubly (Kramers') degenerate. The radiative cascades start with a
recombination of a S-shell e-h pair from the ground states of the
biexciton. The remaining negatively charged exciton (trion) is thus
in an excited state. There are four such states in total. Three
states, in which the two electrons spin wavefunctions are symmetric
under exchange (triplet states) and one in which they are
antisymmetric (singlet).~\cite{Akimov05,Kavokin03,Benny12} The
singlet state marked as $S^*$ in Fig.\ \ref{fig:EnergyDiagram}(a), is higher in energy than the triplet states by the
electron-electron exchange interaction. In the notation we used previously,~\cite{Benny10,Benny11,Benny12} this state is described as follows: $X^{-1}_{S^*}\equiv(1e^12e^1)_S(1h^1)_{\pm 3/2}$ where, $ne^m $ ($nh^m$) denotes $m$ electrons (holes) at the $n$ electronic (hole) state, and the subscript describes the total electronic (hole)  spin.
The degeneracy between the
electronic triplet states $(1e^12e^1)_T$ is further removed by the electron-hole
exchange interaction with the remaining ground state heavy-hole $(1h^1)_{\pm 3/2}$.
Therefore, there are all together 4 doubly degenerate levels of excited trion.
The highest energy states
are these formed from the electronic singlet state $S^*$. Among the levels
formed from the electronic triplet states the lowest one has all
three carriers with parallel spins $(1e^12e^1)_{T_{\pm 1}}(1h^1)_{\pm 3/2}$. These states are dark and cannot
be accessed optically. The triplet level, in which the hole spin is
parallel to the spin of one of the electrons and antiparallel to the
spin of the other electron $(1e^12e^1)_{T_{0}}(1h^1)_{\pm 3/2}$ [the $T_0$ electronic states in Fig.\ \ref{fig:EnergyDiagram}(a)] is next in energy. The highest energy triplet level is that in which the heavy hole spin is antiparallel to the spin
of both electrons $(1e^12e^1)_{T_{\pm 1}}(1h^1)_{\mp 3/2}$ [the $T_{\pm 1}$ electronic states in Fig.\ \ref{fig:EnergyDiagram}(a)]. The figure
presents the radiative (solid arrows) and nonradiative (curly
arrows) spin preserving transitions, by which these levels relax.

In Fig.~\ \ref{fig:EnergyDiagram}(b) we present the measured PL
spectrum from a single QD, optically charged with one electron on
average.~\cite{Benny12} The optical transitions from Fig.\
\ref{fig:EnergyDiagram}(a) are linked to the observed ones by
vertical dashed lines. The main spectral lines are
denoted conventionally, using the initial state of the optical
transition. The spectral lines were identified by their excitation
intensity dependence, their PL excitation spectra,~\cite{Benny12} by
the temporally resolved polarization sensitive intensity correlation
measurements presented below, and by comparison with a many-body
model.~\cite{Poem07} The inset to Fig.\ \ref{fig:EnergyDiagram}(b)
presents horizontally (blue) and vertically (red) linearly polarized PL spectra of the  $T_0$ and $T_1$ charged biexciton and
trion transitions. These four spectral lines are partially linearly
polarized since the e-h anisotropic exchange interaction induces
mixing between the $T_0$ and $T_1$ states.~\cite{Akimov05,Poem10}
We note here in particular that the ratio between the emission
intensities from the two trion lines deviates significantly from the
theoretically predicted ratio~\cite{Akimov05} and from that
experimentally measured previously in negatively~\cite{Akimov05} and
positively~\cite{Akimov05,Poem10} charged QDs. Here, the $T_1$ line
which is predicted to be roughly a factor of two stronger than the
$T_0$ line (as clearly seen in the biexciton lines), is actually
weaker than the trion $T_0$ line. Furthermore, the linewidths of
these spectral lines are significantly larger than the linewidths of
other lines in the PL spectrum.

\begin{figure}
\includegraphics[width=0.48\textwidth]{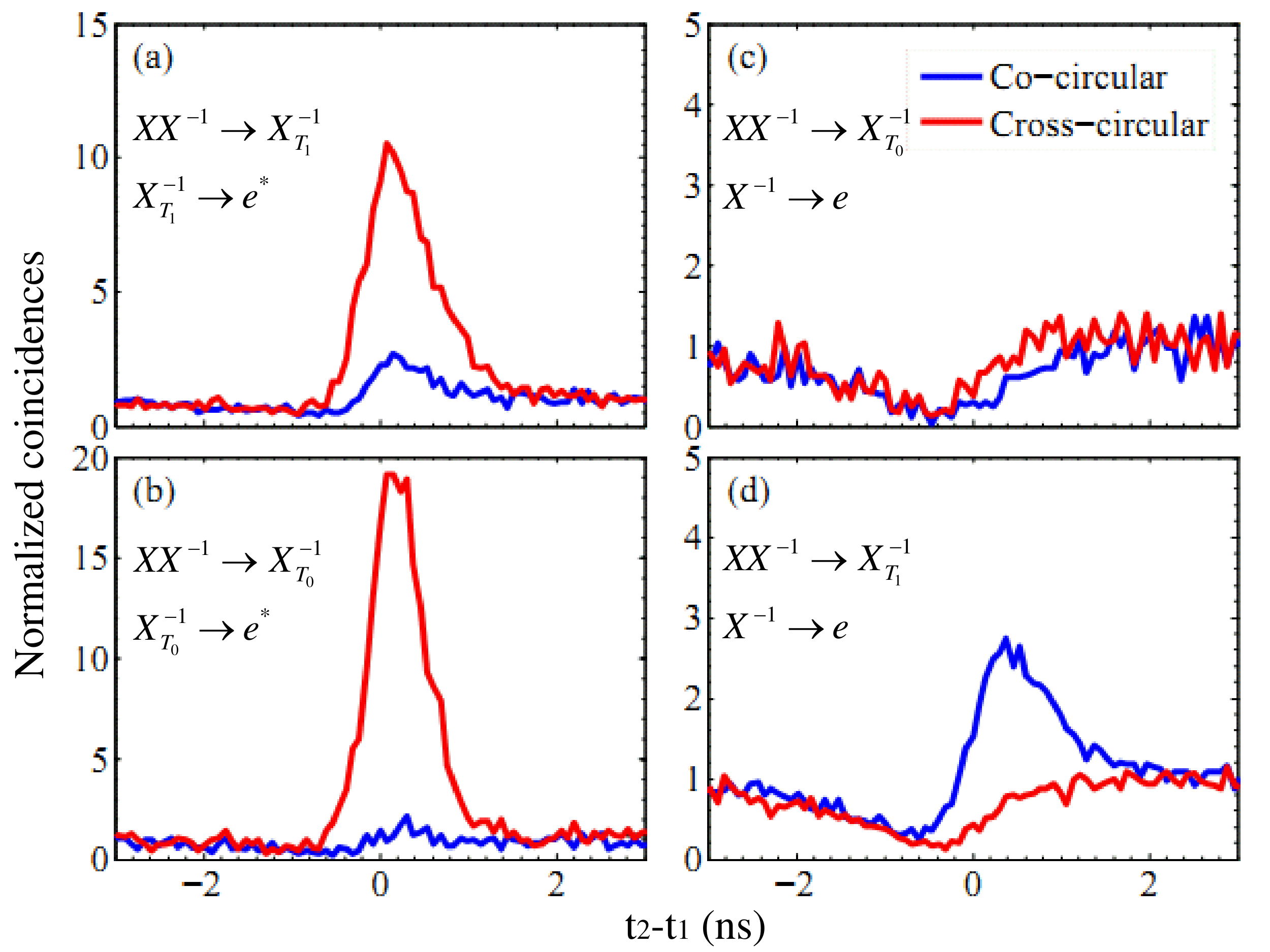}
\caption{\label{fig:TimeResolved} Measured time-resolved
polarization sensitive intensity correlation functions. On the left,
the direct cascades in which $X^{-1}_{T_{1}}$ (a) and
$X^{-1}_{T_{0}}$ (b) are the intermediate levels are presented. On the right, the indirect cascades where $X^{-1}_{T_{1}}$ (c) and
$X^{-1}_{T_{0}}$ (d) decay to the ground singlet trion state $X^{-1}$ are
presented. Blue (red) line stands for measured cross- (co-)
circularly polarized photons.}
\end{figure}

In Fig.\ \ref{fig:TimeResolved} we present polarization sensitive
intensity correlation measurements between photon pairs emitted
during the radiative cascades described in Fig.\
\ref{fig:EnergyDiagram}(a). Red (blue) lines describe co- (cross-)
circularly polarized photons. Figs.\ \ref{fig:TimeResolved}(a-b)
present the results of the spin-conserving cascades:
(a) $XX^{-1}\rightarrow X^{-1}_{T_1}\rightarrow e^*$ and
(b) $XX^{-1}\rightarrow X^{-1}_{T_0}\rightarrow e^*$. These direct
radiative cascades show strong bunching when the two photons are
cross circularly polarized. This clearly demonstrates the expected
spin preserving polarization selection rules. The small bunching
observed when the two photons are co-circularly polarized result
from the fact that the emitted photons are not circularly, but
elliptically polarized, as clearly evidenced by the
partial linear polarization of the four spectral lines [see inset to Fig.\ \ref{fig:EnergyDiagram}(b)].

It is important to note here that the spin allowed direct and
indirect radiative cascades $XX^{-1}\rightarrow
X^{-1}_{S^*}\rightarrow e^*$ and $XX^{-1}\rightarrow
X^{-1}_{S^*}\rightarrow X^{-1}\rightarrow e$, respectively, which were clearly
observed for the case of a singly positively charged
QD,~\cite{Poem10} are not observed here. We believe that the reason for this is the
shorter lifetime of the excited negatively charged singlet trion,
$X^{-1}_{S^*}$. While the positively charged trion decays to its ground $S$ state
within $\sim$25 psec,~\cite{Poem10} the negatively charged excited trion decays to its ground state much faster.
This is due to LO phonon mediated resonant coupling between the ground and excited states.~\cite{Benny10,Benny11}
The strong electron - LO phonon interaction~\cite{Kash85,Kash87} mixes the ground and excited electronic states, facilitating an efficient
electronic relaxation to  the ground state within the lifetime of the LO phonon, which is less than $\sim$7 psec.~\cite{Li99}
Therefore the optical resonance of the negative biexcitonic transition, $XX^{-1}\rightarrow
X^{-1}_{S^*}$, is spectrally broader than that of the positive biexciton, making it more difficult to resolve from the background.
The implications of this resonant coupling of the excited electron to LO phonon are
key issues in understanding the rest of the discussion below.

In Figs.\ \ref{fig:TimeResolved}(c) and \ \ref{fig:TimeResolved}(d) we present intensity
correlation measurements between photons emitted in the biexcitonic
recombinations $T_0$ and $T_{\pm 1}$ and the photon emitted in the
excitonic recombination of the ground state trion $X^{-1}$, respectively. These
indirect radiative cascades require spin blockaded relaxation of
the excited electron in between the two radiative events. Inspecting
Fig.\ \ref{fig:TimeResolved}(c) in which anti-bunching is observed
in both circular polarizations, clearly demonstrates that electronic
spin relaxation from the $T_0$ state to the singlet state is much
slower than the radiative recombination rate. In
contrast, in Fig.\ \ref{fig:TimeResolved}(d), clear and strong
bunching is observed when both photons are co-circularly polarized.
This unambiguously indicates that the $T_1$ state,
$(1e^12e^1)_{T_{\pm 1}}(1h^1)_{\mp 3/2}$, has a very efficient
relaxation channel to the ground singlet state $(1e^2)(1h^{1})_{\pm
3/2}$. The rate of this relaxation, which is faster than the
radiative recombination rate, can be directly extracted from the
measurement, and is found to be $\Gamma_{T_1\rightarrow S}\approx[0.1 \rm{~nsec}]^{-1}$.

This efficient non radiative relaxation path is consistent with the
reduced intensity of the $X^{-1}_{T_{\pm 1}}$ spectral line
observed in  Fig.\ \ref{fig:EnergyDiagram}(b) above.
During this relaxation, both the excited electron and the ground state heavy
hole flip their spins.

We note here that the spin flip-flop mechanism was previously invoked in order to account for negative polarization memory in a single QD~\cite{Cortez02,Golovach08}
and in double QDs.~\cite{Maialle07,Hu11}
Cortez \emph{et al.} for example, suggested  that the electron-hole
exchange interaction, should lead to relaxation accompanied
by spin flip-flop between the electron and the hole.~\cite{Cortez02}
The mixing between the singlet state and the
triplet states which the e-h exchange interaction induces was described by them  in terms of spin flip-flop.
However, since in single self-assembled QDs, the symmetric e-e exchange interaction, which defines the triplet - singlet energy level separation,~\cite{Akimov05} is more than an order of magnitude larger than the e-h exchange interaction ($\Delta_0$), the induced mixing is vanishingly small and cannot possibly lead to the fast relaxation rates that we and others observed experimentally.~\cite{Ware05} B\u{a}descu and Reinecke, therefore, suggested later that a strong asymmetric same-carrier exchange interaction makes the flip-flop mechanism more efficient in some cases.~\cite{Badescu07} Both cases, however,  should have resulted in efficient flip-flop relaxation for positively charged trions as well, in clear contradiction with the experimental observations.~\cite{Poem10}

The rest of this manuscript provides a quantitative theoretical explanation for the efficient spin
flip-flop mechanism that we observe. We show below that the interaction between the electron and an LO phonon
with energy $E_{LO}$, which closely resonates with the energy separation between the single electron levels in the QD ($\Delta E_{1e2e}$),
brings the singlet-triplet levels almost into crossing, thereby, significantly increasing their electron-hole-exchange-induced mixing.

We consider the Hamiltonian
\begin{equation}\label{Ex:Hamiltonian}
\begin{array}{c}
H=H_0+H_{e-h}+H_{e-LO}\\
H_0=H_{carr}+H_{LO}
\end{array}
\end{equation}
where $H_{carr}$ indicates the carrier Hamiltonian which includes the single carrier part of the electrons and hole,  the electron-electron exchange interactions and the electron-hole direct Coulomb interactions, and $H_{LO}$ indicates the single phonon Hamiltonian. The additions to $H_0$ are the electron-hole exchange interactions, $H_{e-h}$,~\cite{Poem07} and the electron-LO phonon interactions, $H_{e-LO}$.~\cite{Stauber00} We consider here the case of the negative trion, as well as that of the neutral exciton. The hole-phonon interaction term is absent from our discussion since the hole in the relevant configurations of the trion and exciton always occupies the same ground energy level. In contrast, the trion and the excited exciton relaxations result from a transition of the electron from its first excited state to its ground state. The energy separation between the first electronic level to the second one, $\Delta E_{1e2e}$, is close to the energy of the LO phonon,
$E_{LO}$, in the semiconductor materials composing the QD~\cite{Hameau99,Benny12} (reported to be in the range $29-36$ meV~\cite{Sarkar05,Heitz99,Lemaitre01,Findeis00}). Therefore, the ground-state trion or exciton with 1 LO phonon are expected to be close in energy to the excited trion or exciton states without an LO phonon, respectively.

The solution to the carriers many-body Hamiltonian in Eq.\ \ref{Ex:Hamiltonian} were discussed in Ref.~\cite{Poem07} Here, however, we choose a more intuitive basis for the subspaces of relevance.
Table \ref{tb:X0} presents the eigenstates and eigenenergies of the Hamiltonian $H_0$ for the case of the neutral exciton. These eigenstates are chosen as the basis for representing the total Hamiltonian. Thus, the matrix in Eq.\ \ref{eq:HexX} represents the electron-hole exchange interactions and electron-phonon Fr\"{o}hlich interaction, $H_{e-h}+H_{e-LO}$, as expressed in the chosen basis.
\begin{table}
\caption{\label{tb:X0} The eigenstates and eigenenergies of the exciton $H_0$ part.}
\begin{ruledtabular}
\begin{tabular}{ll}
Configuration &Energy\\
\hline
$(1h^1)_{+\frac{3}{2}}(2e^1)_{-\frac{1}{2}}(0LO)$&$E^{*}_{carr}$\footnote{$E^{*}_{carr}=E_{1h}+E_{2e}+E_{g}-E_{2e1h1h2e}^{Coul}$, where $E_{ic_1jc_2jc_2ic_1}^{Coul}$ is the energy of the direct Coulomb interaction between carrier $c_1$ (e or h) in state $i$ and carrier $c_2$ in state $j$.}\\
$(1h^1)_{-\frac{3}{2}}(2e^1)_{+\frac{1}{2}}(0LO)$&$E^{*}_{carr}$\\
$(1h^1)_{+\frac{3}{2}}(2e^1)_{+\frac{1}{2}}(0LO)$&$E^{*}_{carr}$\\
$(1h^1)_{-\frac{3}{2}}(2e^1)_{-\frac{1}{2}}(0LO)$&$E^{*}_{carr}$\\
$(1h^1)_{+\frac{3}{2}}(1e^1)_{-\frac{1}{2}}(1LO)$&$E_{carr}+E_{LO}$\footnote{$E_{carr}=E_{1h}+E_{1e}+E_{g}-E_{1e1h1h1e}^{Coul}$.~\cite{Barenco95,Dekel0061}}\\
$(1h^1)_{-\frac{3}{2}}(1e^1)_{+\frac{1}{2}}(1LO)$&$E_{carr}+E_{LO}$\\
$(1h^1)_{+\frac{3}{2}}(1e^1)_{+\frac{1}{2}}(1LO)$&$E_{carr}+E_{LO}$\\
$(1h^1)_{-\frac{3}{2}}(1e^1)_{-\frac{1}{2}}(1LO)$&$E_{carr}+E_{LO}$\\
\end{tabular}
\end{ruledtabular}
\end{table}
\begin{widetext}
\begin{align}
H_{X^0}=
\label{eq:HexX}
\hbox{\scriptsize$
\frac{1}{2}
\left(\begin{array}{cccc|cccc}
\Delta_0^{(1h2e)}&\Delta_1^{(1h2e)}&0&0&2C_{F}&0&0&0\\
\Delta_1^{(1h2e)}&\Delta_0^{(1h2e)}&0&0&0&2C_{F}&0&0\\
0&0&-\Delta_0^{(1h2e)}&\Delta_2^{(1h2e)}&0&0&2C_{F}&0\\
0&0&\Delta_2^{(1h2e)}&-\Delta_0^{(1h2e)}&0&0&0&2C_{F}\\
\hline
2C_{F}&0&0&0&\Delta_0^{(1h1e)}&\Delta_1^{(1h1e)}&0&0\\
0&2C_{F}&0&0&\Delta_1^{(1h1e)}&\Delta_0^{(1h1e)}&0&0\\
0&0&2C_{F}&0&0&0&-\Delta_0^{(1h1e)}&\Delta_2^{(1h1e)}\\
0&0&0&2C_{F}&0&0&\Delta_2^{(1h1e)}&-\Delta_0^{(1h1e)}\\
\end{array}\right)$}.
\end{align}
\end{widetext}

In a similar way,  Table \ref{tb:Trion} presents the eigenstates and eigenenergies of $H_0$ for  the negative trion and Eq.\ \ref{eq:HexTrion} represents the two additional terms to the trion Hamiltonian as expressed in this basis. Here $\widetilde{\Delta}_{0^{\pm}}=\frac{\Delta_0^{(1h1e)}\pm\Delta_0^{(1h2e)}}{2}$, and $\widetilde{\Delta}_{1,2^{\pm}}=\frac{\Delta_{1,2}^{(1h1e)}\pm\Delta_{1,2}^{(1h2e)}}{\sqrt{8}}$.~\cite{Maialle07} $\Delta_{0,1,2}^{(jhie)}$ denote the exchange interaction constants between the hole at level $j$ and the electron at level $i$.~\cite{Ivchenko,Poem10,Poem2010} For clarity, we list in table \ref{tb:params} the parameters discussed above.

\begin{table}
\caption{\label{tb:Trion} The Eigenstates and eigenenergies of the trion $H_0$ part.}
\begin{ruledtabular}
\begin{tabular}{ll}
Configuration &Energy\\
\hline
$(1h^1)_{+\frac{3}{2}}(1e^12e^1)_{T_{+1}}(0LO)$ &   $E_{carr}^*-E^{exch}_{1e2e1e2e}$
\footnote{$E_{carr}^*=E_{1h}+E_{2e}+E_{g}+E^{Coul}_{2e1e1e2e}-E^{Coul}_{1e1h1h1e}-E^{Coul}_{2e1h1h2e}$.~\cite{Barenco95,Dekel0061}}\\

$(1h^1)_{-\frac{3}{2}}(1e^12e^1)_{T_0}(0LO)$    &   $E_{carr}^*-E^{exch}_{1e2e1e2e}$\\

$(1h^1)_{+\frac{3}{2}}(1e^12e^1)_{T_{-1}}(0LO)$ &   $E_{carr}^*-E^{exch}_{1e2e1e2e}$\\

$(1h^1)_{-\frac{3}{2}}(1e^12e^1)_{S^*}(0LO)$    &   $E_{carr}^*+E^{exch}_{1e2e1e2e}$\\

$(1h^1)_{-\frac{3}{2}}(1e^2)_{S}(1LO)$  &   $E_{carr}+E_{LO}$
\footnote{$E_{carr}=E_{1h}+E_{1e}+E_{g}+E^{Coul}_{1e1e1e1e}-2E^{Coul}_{1e1h1h1e}$}\\

$(1h^1)_{-\frac{3}{2}}(1e^12e^1)_{T_{+1}}(0LO)$ &   $E_{carr}^*-E^{exch}_{1e2e1e2e}$\\

$(1h^1)_{+\frac{3}{2}}(1e^12e^1)_{T_0}(0LO)$    &   $E_{carr}^*-E^{exch}_{1e2e1e2e}$\\

$(1h^1)_{-\frac{3}{2}}(1e^12e^1)_{T_{-1}}(0LO)$ &   $E_{carr}^*-E^{exch}_{1e2e1e2e}$\\

$(1h^1)_{+\frac{3}{2}}(1e^12e^1)_{S^*}(0LO)$    &   $E_{carr}^*+E^{exch}_{1e2e1e2e}$\\

$(1h^1)_{+\frac{3}{2}}(1e^2)_{S}(1LO)$  &   $E_{carr}+E_{LO}$\\
\end{tabular}
\end{ruledtabular}
\end{table}

\begin{widetext}
\begin{equation}
H_{X^{-1}}=
\label{eq:HexTrion}
\hbox{\scriptsize$
\left(\begin{array}{ccccc|ccccc}
-\widetilde{\Delta}_{0^{+}}&\widetilde{\Delta}_{2^{+}}&0&-\widetilde{\Delta}_{2^{-}}&0&0&0&0&0&0\\
\widetilde{\Delta}_{2^{+}}&0&\widetilde{\Delta}_{1^+}&\widetilde{\Delta}_{0^-}&0&0&0&0&0&0\\
0&\widetilde{\Delta}_{1^+}&\widetilde{\Delta}_{0^{+}}&\widetilde{\Delta}_{1^-}&0&0&0&0&0&0 \\
-\widetilde{\Delta}_{2^{-}}&\widetilde{\Delta}_{0^-}&\widetilde{\Delta}_{1^-}&0&\sqrt{2}C_{F}&0&0&0&0&0\\
0&0&0&\sqrt{2}C_{F}&0&0&0&0&0&0\\
\hline
0&0&0&0&0&-\widetilde{\Delta}_{0^{+}}&\widetilde{\Delta}_{2^{+}}&0&-\widetilde{\Delta}_{2^{-}}&0\\
0&0&0&0&0&\widetilde{\Delta}_{2^{+}}&0&\widetilde{\Delta}_{1^+}&\widetilde{\Delta}_{0^{-}}&0\\
0&0&0&0&0&0&\widetilde{\Delta}_{1^+}&\widetilde{\Delta}_{0^{+}}&\widetilde{\Delta}_{1^-}&0\\
0&0&0&0&0&-\widetilde{\Delta}_{2^{-}}&\widetilde{\Delta}_{0^-}&\widetilde{\Delta}_{1^-}&0&\sqrt{2}C_{F}\\
0&0&0&0&0&0&0&0&\sqrt{2}C_{F}&0\\
\end{array}\right)$}.
\end{equation}
\end{widetext}

We note that $\Delta_{0\backslash2}$ has contributions mainly from the short range exchange,~\cite{Ivchenko} therefore they hardly depend on the details of the electron or hole spatial wavefunctions, while $\Delta_{1}$ does so.~\cite{Ivchenko,Poem07} Thus, $\widetilde{\Delta}_{0^{-}}$ is negligible. Since in addition $\Delta_2/\Delta_1\ll1$,~\cite{Akimov05,Poem07} it follows that
$\widetilde{\Delta}_{2^{\pm}}$ is rather small.~\cite{Takagahara00,Poem2010} Therefore, the mixing occurs mainly by the $\widetilde{\Delta}_{1^{+}}$ and $\widetilde{\Delta}_{1^{-}}$ terms which couple between the $T_{\pm1}$ and $T_0$ states, and between the $T_{\pm1}$ and $S^*$ states, respectively.

$C_{F}$ represents the Fr\"{o}hlich interaction between the electron and the LO phonon.
Clearly, since $H_{e-ph}$ does not affect the electronic spin, it can only couple ground and excited states of same spin.
This is in perfect agreement with the experiment, where in PL excitation (PLE) spectroscopy of the negative trion, LO phonon associated resonances are observed in the PLE spectrum of the singlet state but not in the spectrum of the triplet states.~\cite{Benny12}

\begin{table}
\caption{\label{tb:params} List of parameters used. }
\begin{ruledtabular}
\begin{tabular}{lp{5cm}p{1.6cm}}
Parameter & Description&Value (meV)\\
\hline
$E_{ic}$  &  The energy of a single carrier (e or h) in level $i$. & -5,  14,   42\footnote{$E_{1h}$,$E_{1e}$,$E_{2e}$, respectively}\\\\
$E_{g}$  &  Band gap of QD material. & 1297\\\\
$E_{ic_1jc_2jc_2ic_1}^{Coul}$  & direct Coulomb interaction between the carrier $c_1$ (e or h) in state $i$ and carrier $c_2$  in state $j$.&22.7,17.0, 24.3, 17.3\footnote{$E_{1e1e1e1e}^{Coul}$,$E_{2e1e1e2e}^{Coul}$, $E_{1e1h1h1e}^{Coul}$,$E_{2e1h1h2e}^{Coul}$, respectively} \\\\
$\Delta E_{1e2e}$ &  The energy difference between the first electronic
level to the second one ($E_{2e}-E_{1e}$).&28 \\\\
$E^{exch}_{1e2e1e2e}$  &   e-e exchange interaction.&5.7 \\\\
$\Delta_{0}^{(jhie)}$  &  The isotropic exchange interaction between h at level $j$ and e at level $i$. Splits between the spin parallel e-h pairs and spin anti parallel pairs. Mostly given by the short-
range interaction.&0.271 \\\\
$\Delta_{1}^{(jhie)}$ &    The anisotropic exchange interaction between h at level $j$ and e at level $i$. Removes the degeneracy between the spin anti parallel e-h pairs. Mostly given by the long-range interaction.&-0.033, 0.324\footnote{$\Delta_{1}^{(1h1e)}$,$\Delta_{1}^{(1h2e)}$, respectively} \\\\
$\Delta_{2}^{(jhie)}$   &   The exchange interaction between h at level $j$ and e at level $i$. Removes the degeneracy  between the spin parallel e-h pairs. Mostly given by the short-range interaction.&-0.0015 \\\\
$\widetilde{\Delta}_{0^{\pm}}$ &   The effective trion exchange term, $\frac{\Delta_0^{(1h1e)}\pm\Delta_0^{(1h2e)}}{2}$&0.2713 \\\\
$\widetilde{\Delta}_{1^{\pm}}$    &The effective trion exchange term, $\frac{\Delta_{1}^{(1h1e)}\pm\Delta_{1}^{(1h2e)}}{\sqrt{8}}$&0.1029, -0.1262 \\\\
$\widetilde{\Delta}_{2^{\pm}}$ &   The effective trion exchange term, $\frac{\Delta_{2}^{(1h1e)}\pm\Delta_{2}^{(1h2e)}}{\sqrt{8}}$&0.0011, 0 \\\\
$E_{LO}$ &  LO phonon energy&32 \\\\
$C_{F}$ &   Fr\"{o}hlich coupling constant&6.4 \\\\
\end{tabular}
\end{ruledtabular}
\end{table}

In order to evaluate the many-body interactions we first model the QD by a 2D parabolic potential for each of the carrier types, as previously described.~\cite{Benny11} The characteristic lengths of the parabolic potentials are chosen to fit the measured energy differences between the first and second single-carrier levels. These permit an almost completely analytical way for calculating the Coulomb and exchange integrals~\cite{Warburton98,Poem07} and the Fr\"{o}hlich energy.~\cite{Stauber00} The standard Fr\"{o}hlich coupling term is evaluated following Stauber \emph{et al.},~\cite{Stauber00} using the renormalized effective dielectric susceptibility, due to the confinement effect.~\cite{Jacak02}
We obtain Fr\"{o}hlich coupling constants of $6.85$ meV for GaAs ($a=0.56$ nm, $\epsilon_{\infty}=10.9$, $\epsilon_0=12.9$,
$E_{LO}
\approx36.6$ meV)  and $6.21$ meV for InAs ($a=0.6$ nm, $\epsilon_{\infty}=12.3$, $\epsilon_0=15.15$,
$E_{LO}
\approx29$ meV).

\begin{figure}
\includegraphics[width=0.48\textwidth]{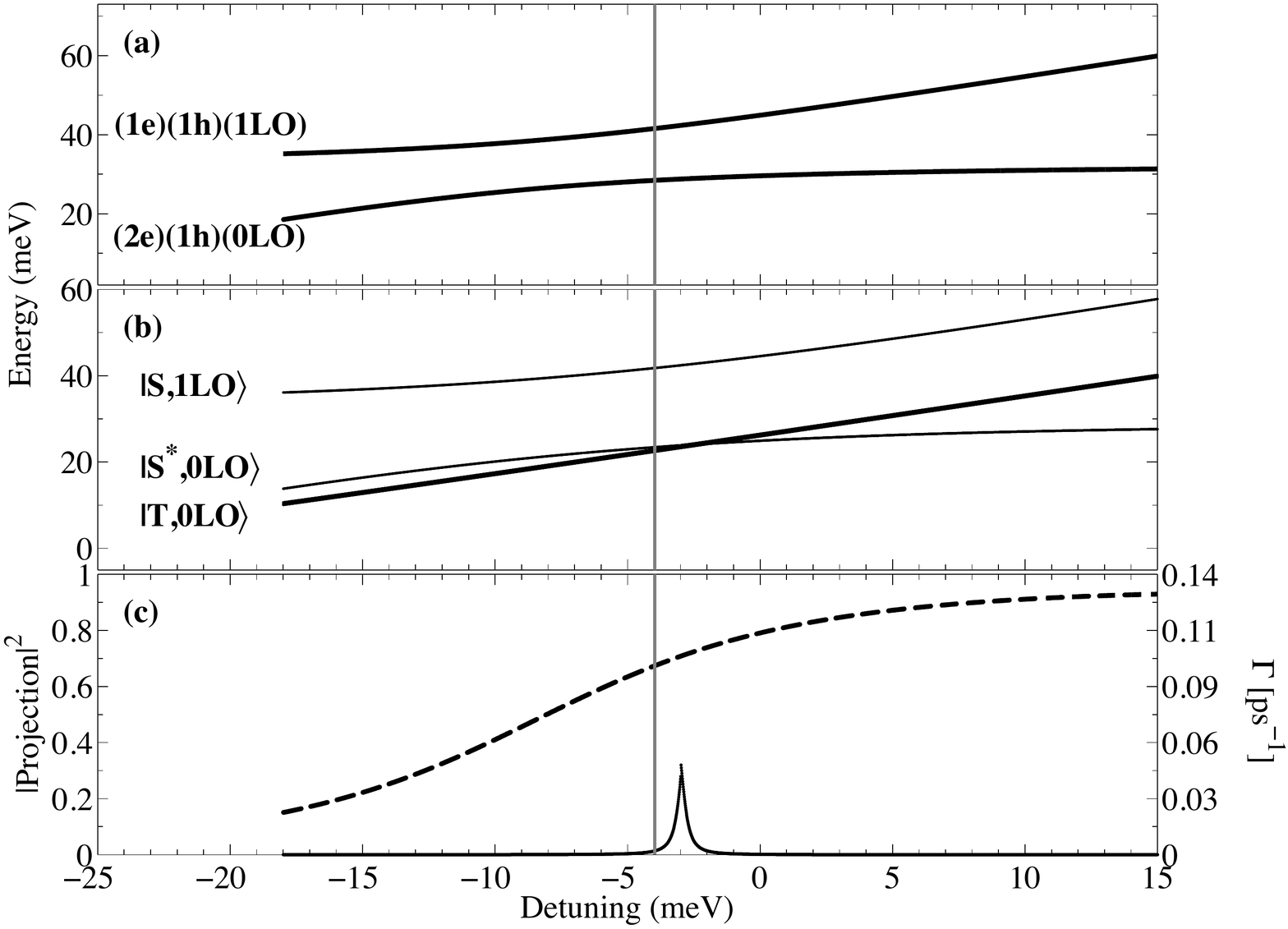}
\caption{\label{fig:AntiCrossing} Calculated energies of the $(2e^1)(1h^1)$ exciton (a) and the negatively charged trion $(1e^12e^1)(1h^1)$ levels (b) as a function of the detuning between the electronic levels separation ($\Delta E_{1e2e}$) and the energy of the LO phonon ($E_{LO}$). Here, $E_{LO} = 32$ meV~\cite{Sarkar05} and the  Fr\"{o}hlich coupling constant is $C_{F}=6.4$ meV (see text). (c) The probability ($|$projection$|^{2}$) of the phonon containing part of the wavefunction [$(1LO)$], in the mixed exciton state, $(2e^1)(1h^1)(0LO)$ (dashed line), and the mixed trion states, $T_{\pm1}$ (solid line), as a function of the detuning (left scale), and the calculated non-radiative decay rates (right scale) of the exciton and trion. We note that the mixed states in (a) and (b) are denoted by their leading terms at negative detuning, where the electron-LO phonon coupling is negligible.}
\end{figure}

Figure \ref{fig:AntiCrossing} presents the calculated eigenenergies of the Hamiltonian of the exciton levels (a) and the trion levels (b) as a function of the detuning between the energy of the phonon,
$E_{LO}$, and the electronic levels separation, $\Delta E_{1e2e}$. Negative detuning means that $\Delta E_{1e2e}$ is smaller than $E_{LO}$. The parameter which we vary in the model in order to achieve different detuning values is the QD lateral area, which mostly affects $\Delta E_{1e2e}$.
Clearly, there is a large difference between the effect of the detuning on the exciton states and its effect on the trion states. In the first case, the $(1e^1)(1h^1)(1LO)$ exciton is mixed with the $(2e^1)(1h^1)(0LO)$ exciton over a rather large range of detunings determined by the magnitude of the Fr\"{o}hlich interaction, $C_F$. In the latter case, the ground singlet $|S,1LO\rangle$ trion is significantly mixed with the excited triplet $|T_{\pm1},0LO\rangle$ trions only when the excited singlet $|S^*,0LO\rangle$ trion is ``pushed" towards the $|T_{\pm1},0LO\rangle$ trion by the Fr\"{o}hlich interaction which mixes the ground and excited singlet trion states. The spin flip-flop interaction becomes important when the energy separation between the excited singlet trion states and the excited triplet states becomes comparable to the $\widetilde{\Delta}_{1^-}$ term, which couples $|S^*,0LO\rangle$ and $|T_{\pm1},0LO\rangle$. Since this term is small, this happens only for a small detuning range.

In Fig.\ \ref{fig:AntiCrossing}(c) we quantitatively evaluate the amount of mixing due to the perturbation Hamiltonian in both the exciton and the trion case (left axis). The exciton curve (dashed line) presents a typical case of two-level mixing. At negative detuning, the lower energy level is mainly composed of the $(2e^1)(1h^1)(0LO)$ state. It includes more and more from the $(1e^1)(1h^1)(1LO)$ state as the detuning diminishes. Then at positive detunings the state becomes mainly $(1e^1)(1h^1)(1LO)$ in nature. For the exciton, this mixing makes the otherwise forbidden optical transition to the excited $(2e^1)(1h^1)$ state allowed, as clearly observed in the PLE spectra of the neutral exciton.~\cite{Benny10}
Moreover, the PLE resonance to the  $(2e^1)(1h^1)$ excited exciton, has a Lorenzian linewidth of about 0.7 meV.~\cite{Benny10} This broad resonance is due to the short electron-LO phonon scattering time ($\sim$0.2 psec~\cite{Kash85}), which couples the excited electron level $(2e^1)$ to the ground one $(1e^1)$. Following absorption, the excited electron rapidly oscillates between its excited and ground levels with exponentially decaying amplitude of oscillations, characterized by the LO phonon lifetime ($\sim$7 psec~\cite{Li99}). The decay rate of the excited exciton is therefore given by dividing  the probability of the phononic part of the wave function (left scale) by the LO phonon lifetime. The result is given on the right scale.

The trion curve (solid line) presents the probability of the phononic state $(1e^2)_{S}(1h^1)_{\pm 3/2}(1LO)$ in the total trion highest energy level, which in addition contains contributions from the $(1e^12e^1)_{T_{\pm 1}}(1h^1)_{\mp 3/2}(0LO)$, $(1e^12e^1)_S(1h^1)_{\pm 3/2}(0LO)$ and $(1e^12e^1)_{T_0}(1h^1)_{\pm 3/2}(0LO)$. At negative detuning the state is mainly composed of the second term ($T_{\pm1}$). The weight of the phonon part in the mixed wavefunction is mostly enhanced in the detuning regime in which the singlet-triplet energy separation is comparable to the energy difference between the mixed triplet states ($T_{\pm1}$ and $T_{0}$). This enhancement is clearly observed in  Fig.\ \ref{fig:AntiCrossing}(c).
The non radiative relaxation rate of the mixed trion state is again calculated by dividing the probability of the phononic part of the wavefunction (left scale) by the LO phonon lifetime ($\sim7$ psec), as expressed by the right scale of  Fig.~\ \ref{fig:AntiCrossing}(c).

In the experimental measurements presented in the inset to Fig.\ \ref{fig:EnergyDiagram}, the intensity of the $T_{\pm 1}$ trion line is roughly 4 times weaker than anticipated, when assuming no non radiative decay rate. This leads to the conclusion that the spin flip-flop time is about 3 times shorter than the radiative recombination time measured to be about $\sim450$ psec.~\cite{Benny10}
This conclusion  well agrees with the measured correlation function in Fig. 2d.  From the deduced magnitude of the flip-flop rate it follows that the detuning is about -4 meV (marked by the vertical line in Fig.\ \ref{fig:AntiCrossing}), which is also consistence with the previously reported measured value of $\Delta E_{1e2e}=27.9$ meV.~\cite{Benny12}

As mentioned above, for simplicity of the previous discussion, we neglected  the difference between $\Delta_0^{(1h1e)}$ and $\Delta_0^{(1h2e)}$. This difference, can be easily incorporated into our calculations. It induces a small mixing between the excited singlet trion state and the triplet $T_{0}$ state. This provides quantitative understanding of the measured small signal in Fig.\ \ref{fig:TimeResolved}(c). There is also a residual mixing between the trion singlet state and the dark trion states. This mixing is so small that it would be significant only for a very small range of detunings.

We note that for positively charged trions, electron-hole spin flip-flop was not observed.~\cite{Poem10} This is easily understood by the much smaller energy separation between the first and second energy levels of the heavy-hole. This in turn leads to a large negative detuning, and thereby to vanishingly small mixing. However, the energy separation between the ground heavy hole level and higher hole energy levels do get comparable to $E_{LO}$.  Indeed, optical transitions that resonate with these hole levels do show typical broadening in PLE spectra of positively charged QDs, evidencing efficient coupling to LO phonons.~\cite{Benny10,Benny11,Benny12}

In summary, we show that in self-assembled quantum dots the Fr\"{o}hlich interaction between electrons and LO phonons may provide an efficient spin flip-flop mechanism for relaxation of excited electrons.
We demonstrated how spin blockaded meta-stable states efficiently relax via spin flip-flop process, resulting from the combined effect of this electron-phonon interaction and the electron-hole exchange interaction.
These processes become important when the electronic energy level separation is comparable to the optical phonon energy. Our quantitative understanding of this phenomenon may provide a novel engineering tool for deterministic spin flip-flop processes in semiconductor nanostructures.


\begin{acknowledgments}
The support of the US-Israel Binational Science Foundation (BSF),
the Israeli Science Foundation (ISF), the Israeli Ministry of
Science and Technology (MOST), and
that of the Technion's RBNI are gratefully acknowledged. K.\ V.\ Kavokin is acknowledged for insightful discussions and  G.\ W.\ Bryant and S.\ Economou for their comments and suggestions.
\end{acknowledgments}


\begin{thebibliography}{41}
\expandafter\ifx\csname natexlab\endcsname\relax\def\natexlab#1{#1}\fi
\expandafter\ifx\csname bibnamefont\endcsname\relax
  \def\bibnamefont#1{#1}\fi
\expandafter\ifx\csname bibfnamefont\endcsname\relax
  \def\bibfnamefont#1{#1}\fi
\expandafter\ifx\csname citenamefont\endcsname\relax
  \def\citenamefont#1{#1}\fi
\expandafter\ifx\csname url\endcsname\relax
  \def\url#1{\texttt{#1}}\fi
\expandafter\ifx\csname urlprefix\endcsname\relax\def\urlprefix{URL }\fi
\providecommand{\bibinfo}[2]{#2}
\providecommand{\eprint}[2][]{\url{#2}}

\bibitem[{\citenamefont{\rm{N.\ Akopian} \textit{et al.}}(2006)}]{Akopian06}
\bibinfo{author}{\bibnamefont{\rm{N.\ Akopian} \textit{et al.}}},
  \bibinfo{journal}{Phys.\ Rev.\ Lett.} \textbf{\bibinfo{volume}{96}},
  \bibinfo{pages}{103501} (\bibinfo{year}{2006}).

\bibitem[{\citenamefont{{Q.\ A.\ Turchette, C.\ J.\ Hood, W.\ Lange, H.\
  Mabuchi, H.\ J.\ Kimble }}(1995)}]{Turchette95}
\bibinfo{author}{\bibnamefont{{Q.\ A.\ Turchette, C.\ J.\ Hood, W.\ Lange, H.\
  Mabuchi, H.\ J.\ Kimble }}}, \bibinfo{journal}{Phys. Rev. Lett.}
  \textbf{\bibinfo{volume}{75}}, \bibinfo{pages}{4710} (\bibinfo{year}{1995}).

\bibitem[{\citenamefont{{E.\ Knill, R.\ Laflamme, G.\ J.\
  Milburn}}(2000)}]{Knill00}
\bibinfo{author}{\bibnamefont{{E.\ Knill, R.\ Laflamme, G.\ J.\ Milburn}}},
  \bibinfo{journal}{Nature} \textbf{\bibinfo{volume}{409}}, \bibinfo{pages}{46}
  (\bibinfo{year}{2000}).

\bibitem[{\citenamefont{{H.\ Kosaka, H.\ Shigyou, Y.\ Mitsumori, Y.\ Rikitake,
  H.\ Imamura, T.\ Kutsuwa, K.\ Arai, K.\ Edamatsu}}(2008)}]{Kosaka08}
\bibinfo{author}{\bibnamefont{{H.\ Kosaka, H.\ Shigyou, Y.\ Mitsumori, Y.\
  Rikitake, H.\ Imamura, T.\ Kutsuwa, K.\ Arai, K.\ Edamatsu}}},
  \bibinfo{journal}{Phys. Rev. Lett.} \textbf{\bibinfo{volume}{100}},
  \bibinfo{pages}{096602} (\bibinfo{year}{2008}).

\bibitem[{\citenamefont{{Y.\ Benny, S.\ Khatsevich, Y.\ Kodriano, E.\ Poem, R.\
  Presman, D.\ Galushko, P.\ M.\ Petroff, D.\ Gershoni}}(2011)}]{Benny10}
\bibinfo{author}{\bibnamefont{{Y.\ Benny, S.\ Khatsevich, Y.\ Kodriano, E.\
  Poem, R.\ Presman, D.\ Galushko, P.\ M.\ Petroff, D.\ Gershoni}}},
  \bibinfo{journal}{Phys. Rev. Lett.} \textbf{\bibinfo{volume}{106}},
  \bibinfo{pages}{040504} (\bibinfo{year}{2011}).

\bibitem[{\citenamefont{{H.\ Kosaka, T.\ Inagaki, Y.\ Rikitake, H.\ Imamura,
  Y.\ Mitsumori, K.\ Edamatsu}}(2008)}]{Kosaka09}
\bibinfo{author}{\bibnamefont{{H.\ Kosaka, T.\ Inagaki, Y.\ Rikitake, H.\
  Imamura, Y.\ Mitsumori, K.\ Edamatsu}}}, \bibinfo{journal}{Nature}
  \textbf{\bibinfo{volume}{457}}, \bibinfo{pages}{702} (\bibinfo{year}{2008}).

\bibitem[{\citenamefont{{D.\ Press, K.\ De Greve, P.\ L.\ McMahon, T.\ D.\
  Ladd, B.\ Friess, C.\ Schneider, M.\ Kamp, S.\ H\"{ö}fling, A.\ Forchel, Y.\
  Yamamoto}}(2010)}]{Press10}
\bibinfo{author}{\bibnamefont{{D.\ Press, K.\ De Greve, P.\ L.\ McMahon, T.\
  D.\ Ladd, B.\ Friess, C.\ Schneider, M.\ Kamp, S.\ H\"{ö}fling, A.\ Forchel,
  Y.\ Yamamoto}}}, \bibinfo{journal}{Nature Photonics}
  \textbf{\bibinfo{volume}{4}}, \bibinfo{pages}{367} (\bibinfo{year}{2010}).

\bibitem[{\citenamefont{{Y.\ Kodriano, I.\ Schwartz, E.\ Poem, Y.\ Benny, R.\
  Presman, T.\ A.\ Truong, P.\ M.\ Petroff, D.\ Gershoni}}(2012)}]{Kodriano12}
\bibinfo{author}{\bibnamefont{{Y.\ Kodriano, I.\ Schwartz, E.\ Poem, Y.\ Benny,
  R.\ Presman, T.\ A.\ Truong, P.\ M.\ Petroff, D.\ Gershoni}}},
  \bibinfo{journal}{Phys. Rev. B} \textbf{\bibinfo{volume}{85}},
  \bibinfo{pages}{241304R} (\bibinfo{year}{2012}).

\bibitem[{\citenamefont{{S.\ Hameau, Y.\ Guldner, O.\ Verzelen, R.\ Ferreira,
  G.\ Bastard, J.\ Zeman, A.\ Lema\^{\i}tre, J.M.
  G\'{e}rard}}(1999{\natexlab{a}})}]{Hameau01}
\bibinfo{author}{\bibnamefont{{S.\ Hameau, Y.\ Guldner, O.\ Verzelen, R.\
  Ferreira, G.\ Bastard, J.\ Zeman, A.\ Lema\^{\i}tre, J.M. G\'{e}rard}}},
  \bibinfo{journal}{Phys.\ Rev.\ Lett.} \textbf{\bibinfo{volume}{83}},
  \bibinfo{pages}{4152} (\bibinfo{year}{1999}{\natexlab{a}}).

\bibitem[{\citenamefont{{S.\ Hameau, Y.\ Guldner, O.\ Verzelen, R.\ Ferreira,
  G.\ Bastard, J.\ Zeman, A.\ Lema\^{\i}tre, J.M.
  G\'{e}rard}}(1999{\natexlab{b}})}]{Hameau99}
\bibinfo{author}{\bibnamefont{{S.\ Hameau, Y.\ Guldner, O.\ Verzelen, R.\
  Ferreira, G.\ Bastard, J.\ Zeman, A.\ Lema\^{\i}tre, J.M. G\'{e}rard}}},
  \bibinfo{journal}{Phys.\ Rev.\ Lett.} \textbf{\bibinfo{volume}{83}},
  \bibinfo{pages}{4152} (\bibinfo{year}{1999}{\natexlab{b}}).

\bibitem[{\citenamefont{{S.\ Hameau, J.\ N.\ Isaia, Y.\ Guldner, E.\ Deleporte,
  O.\ Verzelen, R.\ Ferreira, G.\ Bastard, J.\ Zeman, J. M. G\'{e}rard
  }}(2002)}]{Hameau02}
\bibinfo{author}{\bibnamefont{{S.\ Hameau, J.\ N.\ Isaia, Y.\ Guldner, E.\
  Deleporte, O.\ Verzelen, R.\ Ferreira, G.\ Bastard, J.\ Zeman, J. M.
  G\'{e}rard }}}, \bibinfo{journal}{Phys. Rev. B}
  \textbf{\bibinfo{volume}{65}}, \bibinfo{pages}{085316}
  (\bibinfo{year}{2002}).

\bibitem[{\citenamefont{{T.\ Stauber, R.\ Zimmermann, H.\
  Castella}}(2000)}]{Stauber00}
\bibinfo{author}{\bibnamefont{{T.\ Stauber, R.\ Zimmermann, H.\ Castella}}},
  \bibinfo{journal}{Phys. Rev. B} \textbf{\bibinfo{volume}{62}},
  \bibinfo{pages}{7336} (\bibinfo{year}{2000}).

\bibitem[{\citenamefont{{P.\ Kaczmarkiewicz,P.\
  Machnikowski}}(2010)}]{Kaczmarkiewicz10}
\bibinfo{author}{\bibnamefont{{P.\ Kaczmarkiewicz,P.\ Machnikowski}}},
  \bibinfo{journal}{Phys. Rev. B} \textbf{\bibinfo{volume}{81}},
  \bibinfo{pages}{115317} (\bibinfo{year}{2010}).

\bibitem[{\citenamefont{{L.\ Jacak, J.\ Krasnyj, W.\ Jacak}}(2002)}]{Jacak02}
\bibinfo{author}{\bibnamefont{{L.\ Jacak, J.\ Krasnyj, W.\ Jacak}}},
  \bibinfo{journal}{Phys. Lett. A} \textbf{\bibinfo{volume}{304}},
  \bibinfo{pages}{168} (\bibinfo{year}{2002}).

\bibitem[{\citenamefont{{D.\ V.\ Melnikov, W.\ B.\ Fowler}}(2001)}]{Melnikov01}
\bibinfo{author}{\bibnamefont{{D.\ V.\ Melnikov, W.\ B.\ Fowler}}},
  \bibinfo{journal}{Phys. Rev. B} \textbf{\bibinfo{volume}{64}},
  \bibinfo{pages}{245320} (\bibinfo{year}{2001}).

\bibitem[{\citenamefont{\rm{J. M. Garcia, G. Medeiros-Ribeiro, K. Schmidt, T.
  Ngo, j. L. Feng, A. Lorke, J. Kotthaus, P. M. Petroff}}(1997)}]{Garcia97}
\bibinfo{author}{\bibnamefont{\rm{J. M. Garcia, G. Medeiros-Ribeiro, K.
  Schmidt, T. Ngo, j. L. Feng, A. Lorke, J. Kotthaus, P. M. Petroff}}},
  \bibinfo{journal}{Appl. Phys. Lett.} \textbf{\bibinfo{volume}{71}},
  \bibinfo{pages}{2014} (\bibinfo{year}{1997}).

\bibitem[{\citenamefont{\rm{E.\ Dekel, D.\ V.\ Regelman, D.\ Gershoni, E.\
  Ehrenfreund, W.\ V.\ Schoenfeld, and P.\ M.\ Petroff}}(2001)}]{DekelSSC}
\bibinfo{author}{\bibnamefont{\rm{E.\ Dekel, D.\ V.\ Regelman, D.\ Gershoni,
  E.\ Ehrenfreund, W.\ V.\ Schoenfeld, and P.\ M.\ Petroff}}},
  \bibinfo{journal}{Sol. Stat. Comm.} \textbf{\bibinfo{volume}{117}},
  \bibinfo{pages}{395} (\bibinfo{year}{2001}).

\bibitem[{\citenamefont{{E.\ Poem, Y.\ Kodriano, C.\ Tradonsky, B.\ D.\
  Gerardot, P.\ M.\ Petroff, and D.\ Gershoni}}(2010)}]{Poem10}
\bibinfo{author}{\bibnamefont{{E.\ Poem, Y.\ Kodriano, C.\ Tradonsky, B.\ D.\
  Gerardot, P.\ M.\ Petroff, and D.\ Gershoni}}}, \bibinfo{journal}{Phys.\
  Rev.\ B} \textbf{\bibinfo{volume}{81}}, \bibinfo{pages}{085306}
  (\bibinfo{year}{2010}).

\bibitem[{\citenamefont{{Y.\ Benny, Y.\ Kodriano, E.\ Poem, S.\ Khatsevitch,
  D.\ Gershoni, P.\ M.\ Petroff}}(2011)}]{Benny11}
\bibinfo{author}{\bibnamefont{{Y.\ Benny, Y.\ Kodriano, E.\ Poem, S.\
  Khatsevitch, D.\ Gershoni, P.\ M.\ Petroff}}}, \bibinfo{journal}{Phys. Rev.
  B} \textbf{\bibinfo{volume}{84}}, \bibinfo{pages}{075473}
  (\bibinfo{year}{2011}).

\bibitem[{\citenamefont{\rm{I.\ A.\ Akimov, K.\ V.\ Kavokin, A.\ Hundt, F.\
  Henneberger}}(2005)}]{Akimov05}
\bibinfo{author}{\bibnamefont{\rm{I.\ A.\ Akimov, K.\ V.\ Kavokin, A.\ Hundt,
  F.\ Henneberger}}}, \bibinfo{journal}{Phys. Rev. B.}
  \textbf{\bibinfo{volume}{71}}, \bibinfo{pages}{075326}
  (\bibinfo{year}{2005}).

\bibitem[{\citenamefont{\rm{K.\ V.\ Kavokin}}(2003)}]{Kavokin03}
\bibinfo{author}{\bibnamefont{\rm{K.\ V.\ Kavokin}}}, \bibinfo{journal}{Phys.
  Stat. Sol.} \textbf{\bibinfo{volume}{195}}, \bibinfo{pages}{06157}
  (\bibinfo{year}{2003}).

\bibitem[{\citenamefont{{Y.\ Benny, Y.\ Kodriano, E.\ Poem, D.\ Gershoni, T.\
  A.\ Truong, P.\ M.\ Petroff}}(2012)}]{Benny12}
\bibinfo{author}{\bibnamefont{{Y.\ Benny, Y.\ Kodriano, E.\ Poem, D.\ Gershoni,
  T.\ A.\ Truong, P.\ M.\ Petroff}}}, \bibinfo{journal}{Phys. Rev. B}
  \textbf{\bibinfo{volume}{86}}, \bibinfo{pages}{085306}
  (\bibinfo{year}{2012}).

\bibitem[{\citenamefont{{E.\ Poem, J.\ Shemesh, I.\ Marderfeld, D.\ Galushko,
  N.\ Akopian, D.\ Gershoni, B.\ D.\ Gerardot, A.\ Badolato, P.\ M.\
  Petroff}}(2007)}]{Poem07}
\bibinfo{author}{\bibnamefont{{E.\ Poem, J.\ Shemesh, I.\ Marderfeld, D.\
  Galushko, N.\ Akopian, D.\ Gershoni, B.\ D.\ Gerardot, A.\ Badolato, P.\ M.\
  Petroff}}}, \bibinfo{journal}{Phys.\ Rev.\ B} \textbf{\bibinfo{volume}{76}},
  \bibinfo{pages}{235304} (\bibinfo{year}{2007}).

\bibitem[{\citenamefont{{J.\ A.\ Kash, J.\ C.\ Tsang, J.\ M.\ Hvam
  }}(1985)}]{Kash85}
\bibinfo{author}{\bibnamefont{{J.\ A.\ Kash, J.\ C.\ Tsang, J.\ M.\ Hvam }}},
  \bibinfo{journal}{Phys. Rev. Lett.} \textbf{\bibinfo{volume}{54}},
  \bibinfo{pages}{2151} (\bibinfo{year}{1985}).

\bibitem[{\citenamefont{{J.\ A.\ Kash, S.\ S.\ Jha, J.\ C.\ Tsang
  }}(1987)}]{Kash87}
\bibinfo{author}{\bibnamefont{{J.\ A.\ Kash, S.\ S.\ Jha, J.\ C.\ Tsang }}},
  \bibinfo{journal}{Phys. Rev. Lett.} \textbf{\bibinfo{volume}{58}},
  \bibinfo{pages}{1869} (\bibinfo{year}{1987}).

\bibitem[{\citenamefont{\rm{X.Q. Li, H. Nakayama, Y. Arakawa}}(1999)}]{Li99}
\bibinfo{author}{\bibnamefont{\rm{X.Q. Li, H. Nakayama, Y. Arakawa}}},
  \bibinfo{journal}{Phys. Rev. B} \textbf{\bibinfo{volume}{59}},
  \bibinfo{pages}{5069} (\bibinfo{year}{1999}).

\bibitem[{\citenamefont{{S.\ Cortez, O.\ Krebs, S.\ Laurent, M.\ Senes, X.\
  Marie, P.\ Voisin, R.\ Ferreira, G.\ Bastard}}(2002)}]{Cortez02}
\bibinfo{author}{\bibnamefont{{S.\ Cortez, O.\ Krebs, S.\ Laurent, M.\ Senes,
  X.\ Marie, P.\ Voisin, R.\ Ferreira, G.\ Bastard}}}, \bibinfo{journal}{Phys.
  Rev. Lett.} \textbf{\bibinfo{volume}{89}}, \bibinfo{pages}{207401}
  (\bibinfo{year}{2002}).

\bibitem[{\citenamefont{{V.\ N.\ Golovach, A.\ Khaetskii, D.\
  Loss}}(2008)}]{Golovach08}
\bibinfo{author}{\bibnamefont{{V.\ N.\ Golovach, A.\ Khaetskii, D.\ Loss}}},
  \bibinfo{journal}{Phys. Rev. B} \textbf{\bibinfo{volume}{77}},
  \bibinfo{pages}{045328} (\bibinfo{year}{2008}).

\bibitem[{\citenamefont{{M.\ Z.\ Maialle, M.\ H.\ Degani}}(2007)}]{Maialle07}
\bibinfo{author}{\bibnamefont{{M.\ Z.\ Maialle, M.\ H.\ Degani}}},
  \bibinfo{journal}{Phys. Rev. B} \textbf{\bibinfo{volume}{76}},
  \bibinfo{pages}{115302} (\bibinfo{year}{2007}).

\bibitem[{\citenamefont{{X.\ Hu}}(2011)}]{Hu11}
\bibinfo{author}{\bibnamefont{{X.\ Hu}}}, \bibinfo{journal}{Phys. Rev. B}
  \textbf{\bibinfo{volume}{83}}, \bibinfo{pages}{165322}
  (\bibinfo{year}{2011}).

\bibitem[{\citenamefont{{M.\ E.\ Ware, E.\ A.\ Stinaff, D.\ Gammon, M.\ F.\
  Doty, A.\ S.\ Bracker, D.\ Gershoni, V.\ L.\ Korenev, \c{S}.\ C.\
  B\u{a}descu, Y.\ Lyanda-Geller, T.\ L.\ Reinecke}}(2005)}]{Ware05}
\bibinfo{author}{\bibnamefont{{M.\ E.\ Ware, E.\ A.\ Stinaff, D.\ Gammon, M.\
  F.\ Doty, A.\ S.\ Bracker, D.\ Gershoni, V.\ L.\ Korenev, \c{S}.\ C.\
  B\u{a}descu, Y.\ Lyanda-Geller, T.\ L.\ Reinecke}}}, \bibinfo{journal}{Phys.
  Rev. Lett.} \textbf{\bibinfo{volume}{95}}, \bibinfo{pages}{177403}
  (\bibinfo{year}{2005}).

\bibitem[{\citenamefont{{\c{S}.\ C.\ B\u{a}descu, T.\ L.\
  Reinecke}}(2007)}]{Badescu07}
\bibinfo{author}{\bibnamefont{{\c{S}.\ C.\ B\u{a}descu, T.\ L.\ Reinecke}}},
  \bibinfo{journal}{Phys. Rev. Lett.} \textbf{\bibinfo{volume}{75}},
  \bibinfo{pages}{041309(R)} (\bibinfo{year}{2007}).

\bibitem[{\citenamefont{{D.\ Sarkar, H.\ P.\ van der Meulen, J.\ M.\ Calleja,
  J.\ M.\ Becker, R.\ J.\ Haug, K.\ Pierz}}(2005)}]{Sarkar05}
\bibinfo{author}{\bibnamefont{{D.\ Sarkar, H.\ P.\ van der Meulen, J.\ M.\
  Calleja, J.\ M.\ Becker, R.\ J.\ Haug, K.\ Pierz}}}, \bibinfo{journal}{Phys.
  Rev. B} \textbf{\bibinfo{volume}{71}}, \bibinfo{pages}{081302R}
  (\bibinfo{year}{2005}).

\bibitem[{\citenamefont{{R.\ Heitz,I.\ Mukhametzhanov, O.\ Stier, A.\ Madhukar,
  D.\ Bimberg }}(1999)}]{Heitz99}
\bibinfo{author}{\bibnamefont{{R.\ Heitz,I.\ Mukhametzhanov, O.\ Stier, A.\
  Madhukar, D.\ Bimberg }}}, \bibinfo{journal}{Phys. Rev. Lett.}
  \textbf{\bibinfo{volume}{83}}, \bibinfo{pages}{4654} (\bibinfo{year}{1999}).

\bibitem[{\citenamefont{{A.\ Lema\^{\i}tre, A.\ D.\ Ashmore, J.\ J.\ Finley,
  D.\ J.\ Mowbray, M.\ S.\ Skolnick, M.\ Hopkinson, T.\ F.\
  Krauss}}(2001)}]{Lemaitre01}
\bibinfo{author}{\bibnamefont{{A.\ Lema\^{\i}tre, A.\ D.\ Ashmore, J.\ J.\
  Finley, D.\ J.\ Mowbray, M.\ S.\ Skolnick, M.\ Hopkinson, T.\ F.\ Krauss}}},
  \bibinfo{journal}{Phys.\ Rev.\ B} \textbf{\bibinfo{volume}{63}},
  \bibinfo{pages}{161309} (\bibinfo{year}{2001}).

\bibitem[{\citenamefont{{F.\ Findeis, A.\ Zrenner, G.\ B\"{o}hm, G.\
  Abstreiter}}(2000)}]{Findeis00}
\bibinfo{author}{\bibnamefont{{F.\ Findeis, A.\ Zrenner, G.\ B\"{o}hm, G.\
  Abstreiter}}}, \bibinfo{journal}{Phys.\ Rev.\ B}
  \textbf{\bibinfo{volume}{61}}, \bibinfo{pages}{R10579}
  (\bibinfo{year}{2000}).

\bibitem[{\citenamefont{\rm{A.\ Barenco, M.\ A.\ Dupertuis}}(1995)}]{Barenco95}
\bibinfo{author}{\bibnamefont{\rm{A.\ Barenco, M.\ A.\ Dupertuis}}},
  \bibinfo{journal}{Phys. Rev. B} \textbf{\bibinfo{volume}{52}},
  \bibinfo{pages}{2766} (\bibinfo{year}{1995}).

\bibitem[{\citenamefont{{E.\ Dekel, D.\ Gershoni, E.\ Ehrenfreund, J.\ M.\
  Garcia, P.\ M.\ Petroff}}(2000)}]{Dekel0061}
\bibinfo{author}{\bibnamefont{{E.\ Dekel, D.\ Gershoni, E.\ Ehrenfreund, J.\
  M.\ Garcia, P.\ M.\ Petroff}}}, \bibinfo{journal}{Phys.\ Rev.\ B}
  \textbf{\bibinfo{volume}{61}}, \bibinfo{pages}{11009} (\bibinfo{year}{2000}).

\bibitem[{\citenamefont{{E.\ L.\ Ivchenko, G.\ E.\ Pikus}}(1997)}]{Ivchenko}
\bibinfo{author}{\bibnamefont{{E.\ L.\ Ivchenko, G.\ E.\ Pikus}}},
  \emph{\bibinfo{title}{Superlattices and other Heterostructures}}
  (\bibinfo{publisher}{Springer-Verlag, Berlin}, \bibinfo{year}{1997}).

\bibitem[{\citenamefont{{E.\ Poem, Y.\ .Kodriano, C.\ Tradonsky, N.\ H.\
  Lindner, D.\ Gershoni, B.\ D.\ Gerardot, P.\ M.\ Petroff}}(2010)}]{Poem2010}
\bibinfo{author}{\bibnamefont{{E.\ Poem, Y.\ .Kodriano, C.\ Tradonsky, N.\ H.\
  Lindner, D.\ Gershoni, B.\ D.\ Gerardot, P.\ M.\ Petroff}}}, \bibinfo{journal}{Nature Physics}
  \textbf{\bibinfo{volume}{6}}, \bibinfo{pages}{993} (\bibinfo{year}{2010}).

\bibitem[{\citenamefont{\rm{T.\ Takagahara}}(2000)}]{Takagahara00}
\bibinfo{author}{\bibnamefont{\rm{T.\ Takagahara}}}, \bibinfo{journal}{Phys.\
  Rev.\ B} \textbf{\bibinfo{volume}{62}}, \bibinfo{pages}{16840}
  (\bibinfo{year}{2000}).

\bibitem[{\citenamefont{{R.\ J.\ Warburton, B.\ T.\ Miller, C.\ S.\ D\"{u}rr,
  C.\ B\"{o}defeld, K.\ Karrai, J.\ P.\ Kotthaus, G.\ Medeiros-Ribeiro, P.\ M.\
  Petroff, S.\ Huant}}(1998)}]{Warburton98}
\bibinfo{author}{\bibnamefont{{R.\ J.\ Warburton, B.\ T.\ Miller, C.\ S.\
  D\"{u}rr, C.\ B\"{o}defeld, K.\ Karrai, J.\ P.\ Kotthaus, G.\
  Medeiros-Ribeiro, P.\ M.\ Petroff, S.\ Huant}}}, \bibinfo{journal}{Phys.\
  Rev.\ B} \textbf{\bibinfo{volume}{58}}, \bibinfo{pages}{16221}
  (\bibinfo{year}{1998}).

\end{thebibliography}
\end{document}